\documentclass[aps,pre,twocolumn,showpacs,superscriptaddress,longbibliography]{revtex4-2}
\usepackage{physics, mathtools}
\usepackage{eqnarray, subeqnarray}
\usepackage{amsmath, amssymb, amsthm, amsfonts, bm} % pacotes da AMS para ambientes matemáticos.
\usepackage{xcolor}
\usepackage{fixme, todonotes} % Alerts and notes
\usepackage{verbatim}
\usepackage{scalefnt}
\usepackage[final]{microtype} %melhora a tipografia do texto.
\usepackage{dcolumn}% Align table columns on decimal point
\usepackage{subfigure, float, stfloats, morefloats} % Colocar figuras no local correto usando o [H]
\usepackage{bm}% bold math
\begin{document}
	\preprint{APS/123-QED}
	
	\title{Wound opening in a thin incompressible viscoelastic tissue}% Force line breaks with \\
%	\thanks{Wound opening in a thin incompressible viscoelastic tissue}%
	
	\author{G. M. Carvalho}
	\email{genilson.carvalho@ifc.edu.br}
	\affiliation{Centro de F\'{i}sica Te\'{o}rica e Computacional, Faculdade de Ci\^{e}ncias, Universidade de Lisboa, 1749-016 Lisboa, Portugal}
	\affiliation{Departamento de F\'{\i}sica, Faculdade de Ci\^{e}ncias, Universidade de Lisboa, 1749-016 Lisboa, Portugal}
	\affiliation{Instituto Federal de Educa\c{c}\~{a}o, Ci\^{e}ncia e Tecnologia Catarinense, 89283-064 S\~{a}o Bento do Sul, Santa Catarina, Brasil}
	
	\author{N. A. M. Ara\'ujo}
	\email{nmaraujo@fc.ul.pt}
	\affiliation{Centro de F\'{i}sica Te\'{o}rica e Computacional, Faculdade de Ci\^{e}ncias, Universidade de Lisboa, 1749-016 Lisboa, Portugal}
	\affiliation{Departamento de F\'{\i}sica, Faculdade de Ci\^{e}ncias, Universidade de Lisboa, 1749-016 Lisboa, Portugal}%
	
	\author{P. Patr\'{i}cio}
	\email{pedro.patricio@isel.pt}
	\affiliation{Centro de F\'{i}sica Te\'{o}rica e Computacional, Faculdade de Ci\^{e}ncias, Universidade de Lisboa, 1749-016 Lisboa, Portugal}
	\affiliation{Instituto Superior de Engenharia de Lisboa, Instituto Polit\'{e}cnico de Lisboa, 1959-007 Lisboa, Portugal}%
	
%	\date{\today}% It is always \today, today,
	%  but any date may be explicitly specified
	
	\begin{abstract}
		We develop a model to investigate analytically and numerically the mechanics of wound opening made in a viscoelastic, isotropic, homogeneous, and incompressive thin tissue. This process occurs just immediately after the wound infliction. Before any active biological action has taken place, the tissue relaxes, and the wound opens mostly due to the initial homeostatic tension of the tissue, its elastic and viscous properties, and the existing friction between the tissue and its substrate. We find that for a circular wound the regimes of deformation are defined by a single adimensional parameter $\lambda$, which characterizes the relative importance of viscosity over friction.
	\end{abstract}
	%Use showkeys class option if keyword
	%display desired
	\maketitle
	
	%\tableofcontents
	
	\section{Introduction}	
	Cell and subcellular dynamics, in vivo and in vitro, in response to external stimuli or during the embryonic development of plants and animals, have been the subject of intense research activity during the last decades~\cite{DWeihs_IF_2016,JStephanie_MBEC_2016}. Among the most popular topics is wound healing, which is a process of tissue regeneration for wound closure. This implies collective cellular migration and formation of a contractile cable that connects the cells along the wound edge~\cite{VAjeti_NP_2019,AJacinto_NCB_2001,PMartin_Nature_1992}. Modeling the biophysical mechanisms associated with wound healing is a non trivial challenge~\cite{RTetley_NP_2019,ABrugues_NP_2014,Javierre_JMB_2009,RTranquillo_JSR_1993}.
	
	Recent advances on experimental techniques allow to access the mechanical properties of the tissues and follow their dynamics in real time. These have opened the possibility of studying the process of wound infliction and healing in a systematic and quantitative way~\cite{DSami_WM_2019,SEming_Science_2017,BPurnell_Science_2017,JMReinke_ESR_2012,FHuber_AP_2013}.
	
	Along with the development of experimental techniques, there is a need to develop theoretical and numerical models to shed light on the biochemical and biophysical processes involved in tissue regeneration. There are several ways to model the movement of cells in a tissue~\cite{JLee_BMM_2019,LRoldan_CMAME_2019,AGuerra_JTB_2018,BCamley_JPD_2017,FVermolen_EB_2016,DTartarini_FBB_2016}. The existing models are classified as continuous, particle-based or hybrid models. Continuous models are usually considered to access large length- and time-scales. Particle-based models are appropriate when some level of detail of the particle-particle interaction is of relevance~\cite{DTartarini_FBB_2016,ODea_MTEB_2012}. Hybrid models combine properties of both types~\cite{BCumming_JRSI_2010}. Usually the wound healing models are macroscopic and continuous, being used to investigate the global behavior of cells in the tissue~\cite{JArciero_BJ_2011,JArciero_WRR_2013,LGeris_book_2013}.

	Inspired by recent experimental results~\cite{LCarvalho_JCB_2018}, we propose a theoretical model to investigate analytically and numerically the deformation of epithelial tissues of Drosophila larvae after wound infliction. We focus our study in the first moments of the deformation, the first tens of seconds, before any active biological process comes into play. At this stage, the wound opens under the influence of the initial homeostatic tension of the tissue, its elastic and viscous properties, and the friction between the epithelial tissue and its surroundings. We use a 3D Kelvin-Voigt continuous model, which combines both the elastic and viscous properties of the tissue, and allows the tissue to be initially stretched, resisting to an existing homeostatic tension. The tissue dynamics is given by the Newton's laws in the overdamped regime. By choosing appropriate length and time scales, we find different deformation regimes, which depend on a unique adimensional parameter~$\lambda$, that characterizes the relative importance of the viscosity over friction.
	
	To our knowledge, this adimensional parameter was first introduced in the context of cell mechanics in Ref.~\cite{IBonnet_JRSI_2012}. In this work, the authors severed in vivo the adherens junctions around a disc-shaped domain of  Drosophila pupa dorsal thorax epithelium, comprising typically a hundred cells. They compared the observed deformation of the disk, as it shrunk and relaxed, with the results obtained using a 1D Kelvin-Voigt model to find that the relative importance of viscosity over friction increased with pupa's age (see Fig. 5 of Ref.~\cite{IBonnet_JRSI_2012}). In the context of a Kelvin-Voigt model, this parameter also appears in Ref.~\cite{STlili_EPJE_2015}, where different rheological models are reviewed.

	In the following section, we give the mathematical details of our continuum Kelvin-Voigt model. In the limit of an incompressible thin tissue, we obtain a 2D general equation of motion (Eq. \ref{Eq:motion_2D}). In section III, we solve this equation to obtain the dynamics of the tissue after the infliction of a circular wound. We draw some conclusions in the last section.
	
	\section{Model}
	We model the tissue as an isotropic and homogeneous material, whose mechanical deformation follows the 3D Kelvin-Voigt model. The total stress tensor may be written as \cite{EDill_CRC_2007}:
	\begin{equation}\label{Eq:Tsigma}
		\bm \sigma = 2 G \bm \varepsilon + \lambda_e \Tr(\bm \varepsilon)\bm I+ 
		2 \eta \bm \gamma +\lambda_v \Tr(\bm \gamma)\bm I,
	\end{equation}
	where $\bm \varepsilon$ is the strain tensor, $\bm \gamma$ is the strain-rate tensor, $\bm I$ is the identity tensor, $G$ and $\lambda_e$ are the Lam\'e elastic constants, and $\eta$ and $\lambda_v$ are the dynamic and bulk viscosity coefficients. The first and second (third and fourth) terms of Eq.~\eqref{Eq:Tsigma} correspond to the elastic (viscous) part of the stress tensor, and describe the elastic (viscous) response of each volume element to forces applied tangential and normal to the different surfaces of the element, respectively. In the Kelvin-Voigt model, elastic and viscous stress terms add up.
	
	We use the equilibrium position of the material as the vector coordinate reference $\bm X$. Its deformed position \mbox{${\bm x}({\bm X}, t)$} defines the displacement vector \mbox{${\bm u}({\bm X}, t)={\bm x}(\bm X, t)-{\bm X}.$} For small displacements, the strain tensor and the rate of strain tensor retain only the linear terms in $\bm{u}$:
	\begin{equation}\label{Eq:Tepsilon}
		\bm{\varepsilon} = \frac{1}{2} \left(\bm\nabla \bm{u} + (\bm\nabla \bm{u})^{\mathrm{T}} \right), \quad 
		\bm{\gamma} = \frac{1}{2} \left(\bm\nabla \bm{v} + (\bm\nabla \bm{v})^{\mathrm{T}} \right),
	\end{equation}
	where the differential operator $\bm\nabla$ is defined with respect to the coordinate reference $\bm{X}$, and \mbox{$\bm{v}=\partial \bm{u} / \partial t=\dot{\bm{u}}$}.
	
	The dynamics of the tissue is described by Newton's law of motion:
	\begin{equation}\label{Eq:motion}
		\rho \frac{D \bm{v}}{D t} = \rho \bm{g}+\bm\nabla \cdot \bm{\sigma},
	\end{equation}
	where $\rho$ is the tissue density and $\bm{g}$ the acceleration of gravity. The total time derivative of the velocity is defined as \mbox{$D \bm{v} / D t=\partial \bm{v} / \partial t+\bm{v} \cdot \bm \nabla \bm{v}$}. If the deformation of the tissue is of the order of the size of individual cells, we may neglect the inertial terms and the equation of motion becomes:
	\begin{equation}\label{Eq:motion_simp}
		\bm \nabla \cdot \bm{\sigma} = 0.
	\end{equation}
	
	Let us suppose now that the tissue is also incompressible.
	In this case, we have:
	\begin{equation}\label{Eq:continuity}
		\Tr(\bm \varepsilon)=\Tr(\bm \gamma)=0.
	\end{equation}
	By introducing the pressure field $p$, a Lagrange multiplier which ensures this condition, the stress tensor may be written in a simplified form:
	\begin{equation}\label{Eq:Tsigma_inc}
		\bm \sigma = 2 G \bm \varepsilon + 
		2 \eta \bm \gamma -p \bm I.
	\end{equation}
	
	A thin cellular tissue may be represented as a 2D surface in the plane $x-y$. In this limit, we assume that the normal forces applied to the lower and upper sides of the tissue are much smaller than the longitudinal forces in the bulk. Since the tissue is thin, the normal forces inside the tissue are also negligible and so,
	\begin{equation}
		\sigma_{zz}=0\Leftrightarrow p=2 G\varepsilon_{zz}+2 \eta \gamma_{zz},
	\end{equation}
	everywhere inside the tissue \cite{LLandau_EM_1990}. Taking into account the incompressibility condition, we obtain:
	\begin{equation}
		p=-2 G\Tr(\bm \varepsilon^t)-2 \eta \Tr(\bm \gamma^t), 
	\end{equation}
	where $\bm \varepsilon^t$ and $\bm \gamma^t$ correspond to the 2D strain and strain-rate tensors, defined in the plane \mbox{$x-y$} of the tissue.
	
	Therefore, in the 2D approximation of an incompressible isotropic Kelvin-Voigt tissue, the in-plane 2D stress tensor is given by:
	\begin{equation}\label{Eq:Tsigma_2D}
		\bm \sigma^t= 2 G \left(\bm \varepsilon^t+ \Tr(\bm \varepsilon^t)\bm I^t\right)+ 
		2 \eta\left( \bm \gamma^t+ \Tr(\bm \gamma^t)\bm I^t\right),
	\end{equation}
	where $\bm I^t$ is the identity tensor in 2D.
	
	The friction between the tissue and the substrate is a tangential contact force, exerted on the bottom. Due to the small thickness of the tissue, this force is spread through the interior and it may be described, for the sake of simplicity, as an in-plane force per unit volume \mbox{$\bm f^t = -\zeta \bm v^t$}. The 2D equation of motion becomes:
	\begin{equation}\label{Eq:motion_2D}
		\bm \nabla^t \cdot \bm{\sigma}^t -\zeta \bm v^t = 0.
	\end{equation}
	
	\section{Results}
	\subsection{Tissue under a uniform stress}
	
\begin{figure}[h]
		\centering
		\includegraphics[scale=0.2]{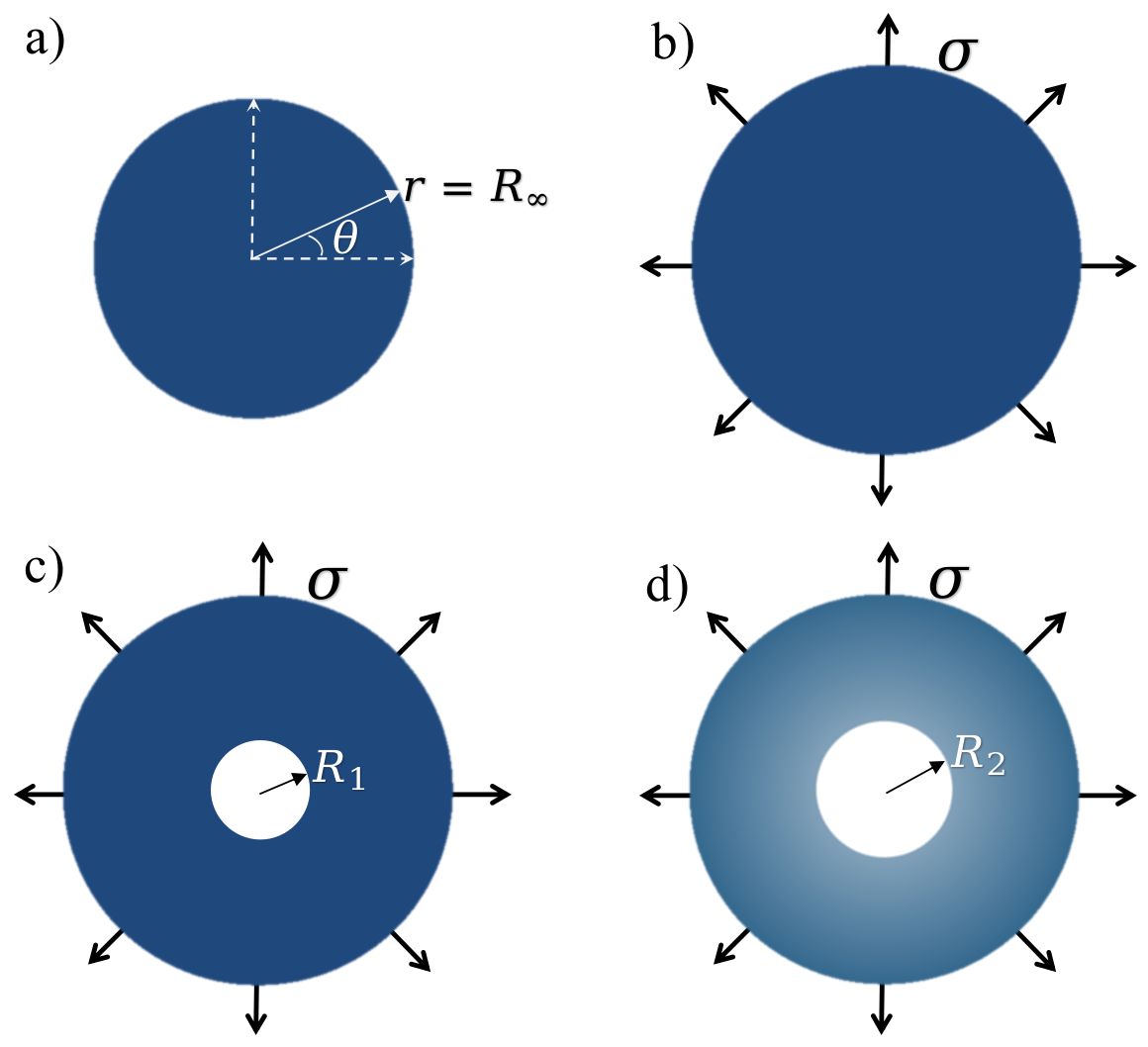}
		\caption{\textbf{a)} A thin circular tissue of radius $R_{\infty}$ at rest. \textbf{b)} Tissue under a uniform radial distribution of forces, where $\sigma$ is the force per unit area, applied at the lateral border of the tissue, leaving it stretched. \textbf{c)} The tissue under tension, just immediately after a circular wound, of radius $R_1$, has been inflicted. There are no forces applied to the internal lateral border of the tissue. \textbf{d)} Final deformation of the tissue, after relaxation. The hole increases to reach a new larger radius $R_2$. The gradation of colors reflects the intensity of the radial tension of the tissue.}
		\label{fig:TC}
	\end{figure}

	We consider first a circular tissue under a uniform distribution of forces, as shown in Fig. \ref{fig:TC}b). At rest, the radius of the circular tissue is $R_{\infty}$ (see Fig. \ref{fig:TC}a). In cylindrical coordinates, the coordinate reference is: \mbox{$\bm{X}^t=(r,\theta)$}, \mbox{$0 \leq r \leq R_{\infty}$} and \mbox{$0 \leq \theta<2 \pi$}. The boundary condition at the periphery of the tissue is
	\begin{equation}\label{Eq:boundary}
		\sigma_{rr}\left(r = R_{\infty}, \theta, t\right) = \sigma.
	\end{equation}
	
	The radial symmetry of the tissue and applied forces suggests a deformation in the form:
	\begin{equation}\label{Eq:deformation}
		\bm{u}^t(r, \theta, t) = u(r, t) \bm{e}_{r}.
	\end{equation}
	The in-plane strain tensor becomes:
	%\begingroup
	%\scalefont{0.9}
	\begin{eqnarray}\label{Eq:cylindrical}
		&&\varepsilon_{rr} = \frac{\partial u}{\partial r}, \quad \varepsilon_{\theta \theta} = \frac{u}{r}, 
		\quad \varepsilon_{r \theta} = 0.
	\end{eqnarray}
	%\endgroup
	The in-plane components of the stress tensor are:
	\begin{eqnarray}\label{Eq:TS_sigma}
		\sigma_{rr} &=& 2 G \left(2 \frac{\partial u}{\partial r} + \frac{u}{r} \right) + 2 \eta \left(2 \frac{\partial \dot{u}}{\partial r} + \frac{\dot{u}}{r} \right), \label{Eq:TS_sigma_rr}\\
		\sigma_{\theta \theta} &=& 2 G \left(\frac{\partial u}{\partial r} + 2 \frac{u}{r} \right) + 2 \eta \left(\frac{\partial \dot{u}}{\partial r} + 2 \frac{\dot{u}}{r} \right), \label{Eq:TS_sigma_theta_theta}\\
		\sigma_{r \theta}&=&0.
	\end{eqnarray}
	
	By symmetry, the polar component of the equation of motion is immediately satisfied. On the other hand, the radial component of the equation of motion \eqref{Eq:motion_2D} is:
	\begin{eqnarray}\label{Eq:motion2}
		\frac{\partial \sigma_{rr}}{\partial r} + \frac{1}{r} \left(\sigma_{rr} - 
		\sigma_{\theta \theta} \right) -\zeta \dot u = 0, \label{Eq:motion2_r}
	\end{eqnarray}
	which, after some algebra, becomes simply
	\begin{equation}\label{Eq:motion2_r2}
		4 \left(G D^{2} u + \eta D^{2} \dot{u} \right) -\zeta \dot u = 0,
	\end{equation}
	where
	\begin{equation}\label{Eq:D2u}
		D^{2} u=\frac{\partial}{\partial r}\left(\frac{1}{r} \frac{\partial}{\partial r}(r u)\right).
	\end{equation}
	
	At equilibrium, the deformation of the tissue must obey the equation $D^{2} u=0$, which has the general solution:
	\begin{equation}\label{Eq:general_sol}
		u(r) = a r + \frac{b}{r},
	\end{equation}
	and the coefficients $a$ and $b$ may be calculated from the boundary conditions. Since $u(r = 0) = 0$ at the center of the tissue, we have $b=0$. The stress is in this case constant at every point of the tissue $\sigma_{rr}=\sigma_{\theta\theta}=6 G a=\sigma$. Thus, we obtain:
	\begin{equation}\label{Eq:general_sol2}
		u(r) = \frac{\sigma r}{6G}.
	\end{equation}
	
	\subsection{Circular wound}
	
	We now consider that a circular wound, of radius $R_1$, is made at the center of the tissue under tension (see Fig.~\ref{fig:TC}c)). Some of the tension is released, and the hole increases to reach a larger radius $R_{2}$ (see Fig.~\ref{fig:TC}d)). The deformation maintains its radial symmetry and the equation of motion is the same (Eq.~\eqref{Eq:motion2_r2}).
	
	In the general case, this equation is solved numerically. However, in the limit of no friction:
	\begin{equation}\label{Eq:motion_viscous_case}
		\left(G D^{2} u + \eta D^{2} \dot{u} \right) = 0,
	\end{equation}
	we may find analytical solutions. In this limit, we have \mbox{$D^2 u=Ae^{-t/\tau}$}, where \mbox{$\tau=\eta/G$} is a relaxation time and $A$ an integration constant. \mbox{$A=0$}, since initially, just immediately after the wound, we have \mbox{$D^2 u=0$}. So, we have the general solution:
	\begin{equation}\label{Eq:general_sol_viscous_case}
		u(r,t) = a(t) r + \frac{b(t)}{r}.
	\end{equation}
	
	The coordinate reference of the wound radius \mbox{$r = R_{0}$} is given by:
	\begin{equation}\label{Eq:R1_R0}
		R_{1} = R_0+u(R_0,t=0)=R_{0} \left(1 + \frac{\sigma}{6G} \right).
	\end{equation}
	We consider that no forces are applied to the tissue at the wound border, so \mbox{$\sigma_{rr}(r = R_{0}, t)=0$}, or:
	\begin{equation}\label{Eq:BC_internal}
		2 G\left(3 a(t) - \frac{b(t)}{R_{0}^{2}}\right) + 2 \eta \left(3 \dot{a}(t) - \frac{\dot{b}(t)}{R_{0}^{2}} \right) = 0.
	\end{equation}
	At the periphery of the tissue, \mbox{$\sigma_{rr}(r = R_\infty, t)=\sigma$}:
	\begin{equation}\label{Eq:BC_external}
		2 G \left(3 a(t) - \frac{b(t)}{R_{\infty}^{2}} \right) + 2 \eta\left(3 \dot{a}(t) - \frac{\dot{b}(t)}{R_{\infty}^{2}}\right) = \sigma.
	\end{equation}
	If \mbox{$R_0\ll R_\infty$}, the final set of equations for $a$ and $b$ may then be written in the form:
	\begin{eqnarray}
		\label{Eq:Diff_eq_a}
		a+\tau\dot a=\frac{\sigma}{6G} \left( \frac{R^2_\infty}{R^2_\infty-R^2_0}\right)\approx \frac{\sigma}{6G} ,\\
		b+\tau\dot b=\frac{\sigma R^2_0}{2G}\left( \frac{R^2_\infty}{R^2_\infty-R^2_0}\right) \approx \frac{\sigma R^2_0}{2G}.
		\label{Eq:Diff_eq_b}
	\end{eqnarray}
	
	Using the initial conditions:
	\begin{equation}\label{Eq:a_and_b}
		a(t=0) = \frac{\sigma}{6G}, \quad b(t=0) = 0.
	\end{equation}
	we obtain:
	%
	% The final result without the approximation R_0<<R_infty
	%
	%\begin{eqnarray}
	%\label{Eq:DeformationHoleA}
	%a(t)&=&\frac{\sigma}{6G}\left(\frac{R^2_\infty}{R^2_\infty-R^2_0}\right)\left(1-\frac{R_0^2}{R_\infty^2}e^{-t/\tau}\right),\\
	%b(t)&=&\frac{\sigma R_0^2}{2G}\left(\frac{R^2_\infty}{R^2_\infty-R^2_0}\right)\left(1-e^{-t/\tau}\right).
	%\label{Eq:DeformationHoleB}
	%\end{eqnarray}
	%or, if $R_0\ll R_\infty$: 
	\begin{eqnarray}
		\label{Eq:DeformationHoleC}
		a(t)&\approx&\frac{\sigma}{6G},\\
		b(t)&\approx&\frac{\sigma R_0^2}{2G}\left(1-e^{-t/\tau}\right).
		\label{Eq:DeformationHoleD}
	\end{eqnarray}
	
	In particular, we may now determine the time evolution of the wound opening, \mbox{$R(t)= R_{0} + u\left(R_{0}, t \right)$}. In the limit of small displacements, to linear order in $\sigma/G$, we have:
	\begin{eqnarray}
		\label{Eq:R(t)_R1}
		R(t) \approx R_1\left[1+\frac{\sigma}{2G}\left(1-e^{-t/\tau}\right) \right].
	\end{eqnarray}
	
	In the presence of friction, the deformation of the tissue under tension is different. Nevertheless, their equilibrium initial and final states is the same, as the velocities of these configurations are zero. To obtain the time dependence of the deformation, we must solve this equation numerically.
		
	If we choose \mbox{$L = R_0$} and \mbox{$T = \tau$} as unit length and time scales, the adimensional equation of motion is then:
	\begin{equation}\label{Eq:deformation_adimen}
		D^{2} u + D^{2} \dot{u} - \frac{\dot{u}}{\lambda^2} = 0,
	\end{equation}
	where \mbox{$\lambda = L_{\eta} / R_{0}$}, and \mbox{$L_{\eta} =2 \sqrt{\eta / \zeta}$} is the {\sl viscous} length. The parameter $\lambda$ sets the possible regimes of deformation. When \mbox{$\lambda\to\infty$}, we recover the limit where friction is negligible, discussed above.
	
	Numerically, it is convenient to use \mbox{$T=\tau/\lambda^2$} as the time unit. This choice yields the following equation of motion:
	\begin{equation}\label{Eq:D_deformation_adimen}
		D^{2} u + \lambda ^2D^{2} \dot{u} -\dot{u} = 0.
	\end{equation}
	If \mbox{$\lambda\to 0$}, we may use an explicit method to integrate this equation. It will be stable if we choose a sufficiently small time step. However, as $\lambda$ increases, the method rapidly gets unstable. To solve this equation for every choice of $\lambda$, we use the implicit method described in the appendix, with second order precision in space and time. The boundary conditions are expressed through the stress tensor. The stress tensor is adimensionalized by dividing it by the elastic modulus $G$. Using this latter choice for unit space and time scales, we have:
	\begin{eqnarray}\label{Eq:bc_adimen}
		\sigma_{r r}= D u+\lambda^{2} D \dot{u},
	\end{eqnarray}
	where
	\begin{equation}\label{Eq:sigma_rr_adimen}
		D = 2\left(2 \frac{\partial}{\partial r}+\frac{1}{r}\right).
	\end{equation}
	So, the explicit adimensionalized boundary conditions are \mbox{$\sigma_{rr}(r=1)=0$} and \mbox{$\sigma_{rr}(r=R_\infty/R_0)=\sigma$}.
	
\begin{figure}[h]
		\centering
		\includegraphics[scale=1.12]{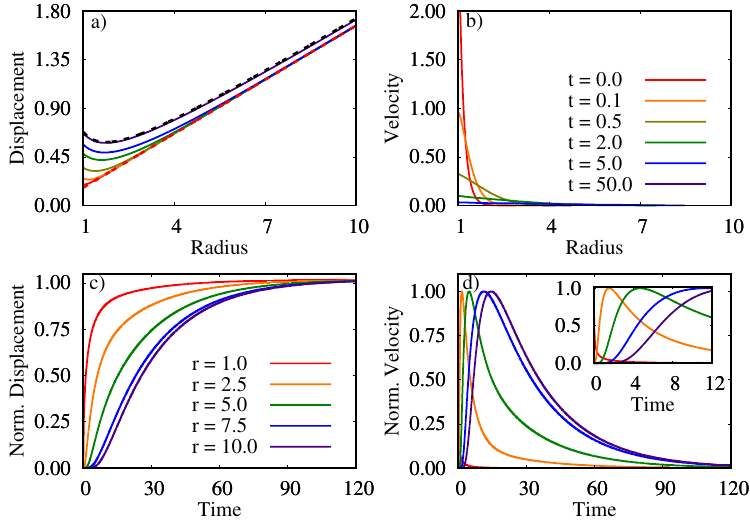}
		\caption{Tissue deformation after wound infliction, 
			for $\lambda=0.1$, $R_\infty=10$ and $\sigma=1$ (friction regime).
			a) Displacement $u(r,t)$ {\it vs} radius $r$, for different instants $t$ of the deformation. The dashed red and black lines represent respectively the displacement before wound infliction ($t=0$), and at the final relaxed opened state of the wound ($t=\infty$);
			b) Velocity \mbox{$v(r,t)=\dot u(r,t)$} {\it vs} radius $r$, for different instants $t$ of the deformation;
			c) Normalized displacement (\mbox{$(u - u_{i}) / (u_{f} - u_{i})$}, with $u_{i}$ the initial displacement and $u_{f}$ the final displacement) {\it vs} time, in different positions $r$ of the tissue.
			d) Normalized velocity (\mbox{$v / v_{\max}$, with $v_{max}$} the maximum velocity attained at a particular position) {\it vs} time, in different positions $r$ of the tissue.}
		\label{fig:lbd_1e-2}
	\end{figure}

\begin{figure}[h]
		\centering
		\includegraphics[scale=1.12]{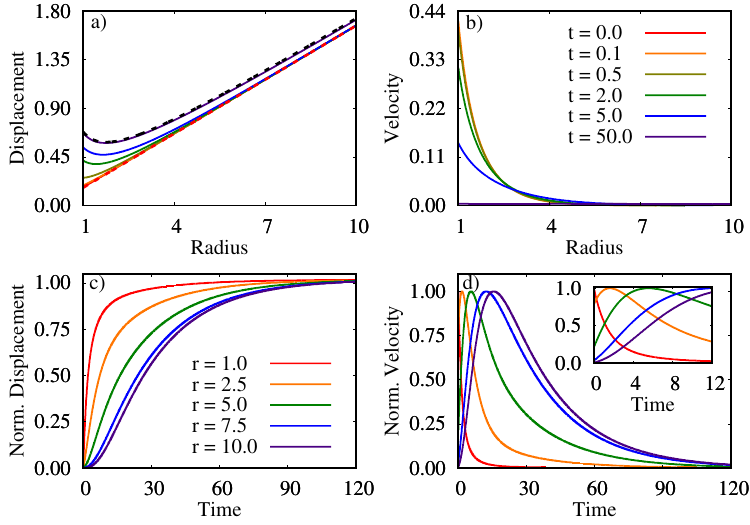}
		\caption{The same as in Fig.~\ref{fig:lbd_1e-2}, but for $\lambda = 1$ (intermediate regime).}
		\label{fig:lbd_1e0}
	\end{figure}

\begin{figure}[h]
		\centering
		\includegraphics[scale=1.12]{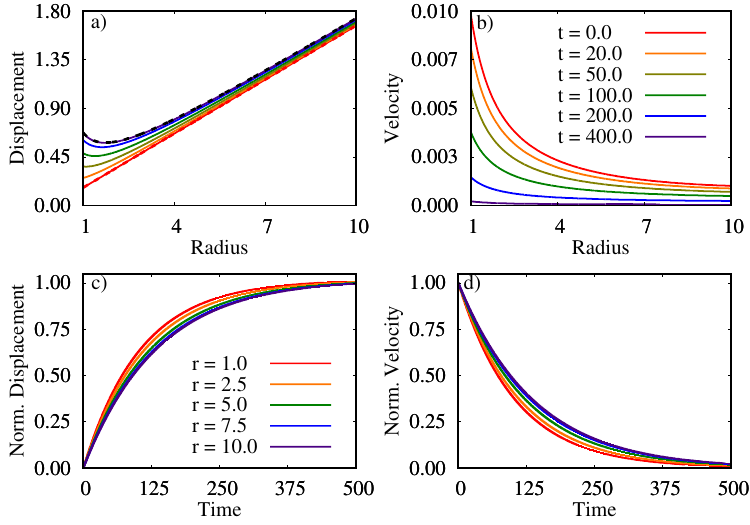}
		\caption{The same as in Fig.~\ref{fig:lbd_1e-2}, but with $\lambda = 10$ (viscous regime).}
		\label{fig:lbd_1e2}
	\end{figure}

	Figures~\ref{fig:lbd_1e-2}, \ref{fig:lbd_1e0} and \ref{fig:lbd_1e2} show the results for the displacement $u(r,t)$ and the velocity \mbox{$v(r,t)=\dot u(r,t)$}, for three different values of \mbox{$\lambda=0.1,1,10$} (corresponding respectively to the friction, intermediate and viscous regimes), \mbox{$R_\infty=10$} and \mbox{$\sigma=1$}.
	
	The panel a) of each figure shows the displacement as a function of the radius, for different instants $t$. The dashed red and black lines represent respectively the displacement before wound infliction ($t=0$), and at the final relaxed opened state of the wound ($t=\infty$). These representations show already the differences between the three different deformation regimes. However these differences stand out in the other plots, as we discuss in what follows.

	The panels b) show the velocity as a function of the radius, also for different instants $t$ it is clear that the velocity is initially (at $t=0$) higher closer the wound. The velocity decreases rapidly as we move away from the wound. In fact, the characteristic length of this decay is of the order of the viscous length \mbox{$L_\eta=2\sqrt{\eta/\zeta}=\lambda R_0$}. In the friction regime (Fig.~\ref{fig:lbd_1e-2}~b)), the decay length is very small ($L_\eta=0.1$), and the initial velocity is peaked at the border of the wound. In the viscous regime (Fig.~\ref{fig:lbd_1e2}~b)), the decay length is large ($L_\eta=10$), of the same size as the tissue itself. The inital velocity decay length is then a signature of the regime, and may be of experimental interest to access the relative importance of the viscosity over friction.
	
	Panels c) show the time dependence of the normalized displacement \mbox{$(u - u_{i}) / (u_{f} - u_{i})$}, with $u_{i}$ the initial displacement and $u_{f}$ the final displacement, for different positions $r$. With these figures, it is possible to understand that in friction regimes (small $\lambda$), the wound opening occurs faster near the wound ($r=1$), whereas away from the wound ($r>7.5$), the tissue stays almost immobile for the initial times, before relaxing to its final state, with a slight different relaxation time. In viscous regimes (large $\lambda$), the tissue globally relaxes with a characteristic time which depends solely on the ratio between its viscous and elastic properties.
	
	Panels d) show, for different positions $r$, the time dependence of the normalized velocity \mbox{$v / v_{\max}$}, with $v_{max}$ the maximum velocity attained at a particular position. From these figures, it is clear that in friction regimes, the positions away from the wound do not move immediately. Instead, their speeds increase up to a certain value, and only afterwards start to decrease until they achieve their relaxed states. The results suggest that there is a propagation wave affecting the maximum normalized velocity for each position. This effect is however hardly noticed, because the velocities far from the wound are already very small. In the viscous regimes, the velocity profile is almost the same at each point of the tissue: the tissue feels almost instantaneously the wound infliction everywhere.
	
	\section{Final Remarks}
	We developed a theoretical model to investigate analytically and numerically the mechanical deformation of an epithelial tissue after a circular wound has been inflicted. 
	The tissue was described as a continuous isotropic, homogeneous and incompressible thin material, obeying the 3D Kelvin-Voigt model. This model takes into account the elastic and viscous properties of the tissue, and allows the tissue to be stretched at equilibrium, under the effect of a homeostatic pressure. Friction between the tissue and its surroundings was also considered. We determined the passive mechanical response of the tissue, after wound infliction, without considering any active biological effects which will try to close the wound, and heal the tissue. This behavior is consistent with the passive physical behaviour observed for the first tens of seconds for the Drosophila larvae studied in Ref.~\cite{LCarvalho_JCB_2018}.
	
	By choosing appropriate length and time scales, we found different deformation regimes, depending on a unique adimensional parameter $\lambda$, which characterizes the relative importance of the viscosity $\eta$ over friction $\zeta$. Although the final relaxed state is the same, for all values of the viscosity or friction, the dynamics of the deformation presents distinct features. In friction regimes, for small $\lambda$, the deformation is initially concentrated at the border of the wound, whereas in viscous regimes, for large $\lambda$, the deformation evolves globally. In fact, if only viscosity is present everywhere, and only depends friction is negligeable, the normalized displacement time evolution is exactly the same, everywhere, and only depends on a relaxation time simply defined by the ratio between the viscous and elastic properties of the tissue.
	
	The experimental characterization of these different regimes may be accessed through the initial velocity space profile, \mbox{$v(r,t=0)$}. The initial velocity field typically decays from the wound boundary in a typical length given
	by \mbox{$L_\eta=\lambda R_0=2\sqrt{\eta/\zeta}$}, which relates viscosity and friction.
	
	\appendix*
	\section{Numerical integration}
	To integrate the equations numerically, we discretize space and time:
	\begin{eqnarray}
		\label{Eq:wound_radius}
		r_{i} &=& 1 + (i-1) \Delta r, \quad (i = 1, ..., N), \\
		t_{n} &=& n \Delta t, \quad (n=0,1, ...),
	\end{eqnarray}
	with \mbox{$\Delta r = \left(R_{\infty}-1\right) /(N-1)$}. We write \mbox{$u\left(r_{i}, t_{n}\right) = u_{i}^{n}$}, for simplicity. The discretized adimensional equation of motion (Eq.~\eqref{Eq:D_deformation_adimen}) is approximated by:
	\begin{equation}\label{Eq:motion_discretized}
		D^{2} u_{i}^{n+1} + \lambda^{2} \frac{D^{2} u_{i}^{n+1}-D^{2} u_{i}^{n-1}}{2 \Delta t} - \frac{u_{i}^{n+1}-u_{i}^{n-1}}{2 \Delta t} = 0
	\end{equation}
	with
	\begin{equation}\label{Eq:D2u_i}
		D^{2} u_{i} = \frac{u_{i+1} - 2 u_{i} + u_{i-1}}{\Delta r^{2}} + \frac{1}{r_{i}} \frac{u_{i+1} - u_{i-1}}{2 \Delta r} - \frac{1}{r_{i}^{2}} u_{i}.
	\end{equation}
	If we use matrix notation, we may write the \mbox{$N-2$} equations, with \mbox{$i = 2, ..., N-1$}:
	\begin{equation}\label{Eq:matrixial_notation}
		\left(-\delta_{i j} + \left(\lambda^{2} + 2 \Delta t\right) D_{i j}^{2} \right) u_{j}^{n+1} = \left(-\delta_{i j} + \lambda^{2} D_{i j}^{2}\right)u_{j}^{n-1},
	\end{equation}
	where we used Einstein’s convention for the sum of repeated indexes, and
	\begin{eqnarray}\label{Eq:Einstein_convention}
		D_{i, i-1}^{2} &=& \frac{1}{\Delta r^{2}} - \frac{1}{2 r_{i} \Delta r} \label{Eq:D2_i-1}\\
		D_{i, i}^{2} &=& -\frac{2}{\Delta r^{2}}-\frac{1}{r_{i}^{2}} \label{Eq:D2_i}\\
		D_{i, i+1}^{2} &=& \frac{1}{\Delta r^{2}}+\frac{1}{2 r_{i} \Delta r} \label{Eq:D2_i+1}
	\end{eqnarray}
	and \mbox{$D_{ij}^{2} = 0$} otherwise.
	
	The other two equations are given from the boundary conditions. For \mbox{$r_{1} = 1$}, we have \mbox{$\sigma_{rr}=0$}.
	The discretized adimensional boundary condition becomes:
	\begin{equation}\label{Eq:discretization_2}
		D u_{i}^{n+1} + \lambda^{2} \frac{D u_{i}^{n+1}-D u_{i}^{n-1}}{2 \Delta t} = 0
	\end{equation}
	or in matricial notation:
	\begin{equation}\label{Eq:matricial_notation2}
		\left(\lambda^{2} + 2 \Delta t\right) D_{1 j} u_{j}^{n+1} = \lambda^{2} D_{1 j} u_{j}^{n-1},
	\end{equation}
	with
	\begin{eqnarray}
		D_{11} &=& -\frac{4}{\Delta r} + \frac{2}{r_{1}}, \label{Eq:D11}\\
		D_{12} &=& \frac{4}{\Delta r}, \label{Eq:D12}
	\end{eqnarray}
	and \mbox{$D_{1j} = 0$} for all other values of $j$. For \mbox{$r_{N} = R_{\infty}$}, \mbox{$\sigma_{rr}=\sigma$}.
	The discretized adimensional boundary condition becomes:
	\begin{equation}\label{Eq:BC_external_adimen_matricial}
		\left(\lambda^{2}+2 \Delta t\right) D_{N j} u_{j}^{n+1} = \lambda^{2} D_{N j} u_{j}^{n-1} + 2 \Delta t \sigma,
	\end{equation}
	with
	\begin{eqnarray}
		D_{N, N-1} &=& -\frac{4}{\Delta r}, \label{Eq:D_N-1}\\
		D_{N, N} &=& \frac{4}{\Delta r}+\frac{2}{r_{N}}, \label{Eq:D_N}
	\end{eqnarray}
	and \mbox{$D_{Nj} = 0$} for all other values of $j$.

	In sum, we have a matricial equation of the kind:
	\begin{equation}\label{Eq:matricial_equation}
		A_{i j} u_{j}^{n+1} = B_{i j} u_{j}^{n-1}+C_{i},
	\end{equation}
	where the matrix $A_{ij}$ is
	\begingroup
	\scalefont{0.8}
	\begin{eqnarray}
		A_{1 j} &=& \left(\lambda^{2}+2 \Delta t\right) D_{1 j}, \label{Eq:matrix_A1j} \\
		A_{i j} &=& -\delta_{i j}+\left(\lambda^{2}+2 \Delta t\right) D_{i j}^{2}, \quad(i=2, \ldots, N-1), \label{Eq:matrix_Aij}\\
		A_{N j} &=& \left(\lambda^{2}+2 \Delta t\right) D_{N j}, \label{Eq:matrix_ANj}
	\end{eqnarray}
	\endgroup
	the matrix $B_{ij}$ is
	\begin{eqnarray}
		B_{1 j} &=& \lambda^{2} D_{1 j}, \label{Eq:matrix_B1j} \\
		B_{i j} &=& -\delta_{i j}+\lambda^{2} D_{i j}^{2}, \quad(i=2, \ldots, N-1), \label{Eq:matrix_Bij} \\
		B_{N j} &=& \lambda^{2} D_{N j}, \label{Eq:matrix_BNj}
	\end{eqnarray}
	and the vector $C_{i}$ is given by \mbox{$C_{N} = 2\Delta t \sigma$} and \mbox{$C_{i} = 0$} otherwise.

	\bibliographystyle{apsrev} %APS Style
	\bibliography{wound_formation.bib}

\begin{thebibliography}{30}
\expandafter\ifx\csname natexlab\endcsname\relax\def\natexlab#1{#1}\fi
\expandafter\ifx\csname bibnamefont\endcsname\relax
  \def\bibnamefont#1{#1}\fi
\expandafter\ifx\csname bibfnamefont\endcsname\relax
  \def\bibfnamefont#1{#1}\fi
\expandafter\ifx\csname citenamefont\endcsname\relax
  \def\citenamefont#1{#1}\fi
\expandafter\ifx\csname url\endcsname\relax
  \def\url#1{\texttt{#1}}\fi
\expandafter\ifx\csname urlprefix\endcsname\relax\def\urlprefix{URL }\fi
\providecommand{\bibinfo}[2]{#2}
\providecommand{\eprint}[2][]{\url{#2}}

\bibitem[{\citenamefont{Weihs et~al.}(2016)\citenamefont{Weihs, Gefen, and
  Vermolen}}]{DWeihs_IF_2016}
\bibinfo{author}{\bibfnamefont{D.}~\bibnamefont{Weihs}},
  \bibinfo{author}{\bibfnamefont{A.}~\bibnamefont{Gefen}}, \bibnamefont{and}
  \bibinfo{author}{\bibfnamefont{F.~J.} \bibnamefont{Vermolen}},
  \bibinfo{journal}{Interface Focus} \textbf{\bibinfo{volume}{6}},
  \bibinfo{pages}{20160038} (\bibinfo{year}{2016}).

\bibitem[{\citenamefont{Jorgensen and Sanders}(2016)}]{JStephanie_MBEC_2016}
\bibinfo{author}{\bibfnamefont{S.~N.} \bibnamefont{Jorgensen}}
  \bibnamefont{and} \bibinfo{author}{\bibfnamefont{J.~R.}
  \bibnamefont{Sanders}}, \bibinfo{journal}{Med. Biol. Eng. Comput.}
  \textbf{\bibinfo{volume}{54}}, \bibinfo{pages}{1297} (\bibinfo{year}{2016}).

\bibitem[{\citenamefont{Ajeti et~al.}(2019)\citenamefont{Ajeti, Tabatabai,
  Fleszar, Staddon, Seara, Suarez, Yousafzai, Bi, Kovar, Banerjee
  et~al.}}]{VAjeti_NP_2019}
\bibinfo{author}{\bibfnamefont{V.}~\bibnamefont{Ajeti}},
  \bibinfo{author}{\bibfnamefont{A.~P.} \bibnamefont{Tabatabai}},
  \bibinfo{author}{\bibfnamefont{A.~J.} \bibnamefont{Fleszar}},
  \bibinfo{author}{\bibfnamefont{M.~F.} \bibnamefont{Staddon}},
  \bibinfo{author}{\bibfnamefont{D.~S.} \bibnamefont{Seara}},
  \bibinfo{author}{\bibfnamefont{C.}~\bibnamefont{Suarez}},
  \bibinfo{author}{\bibfnamefont{M.~S.} \bibnamefont{Yousafzai}},
  \bibinfo{author}{\bibfnamefont{D.}~\bibnamefont{Bi}},
  \bibinfo{author}{\bibfnamefont{D.~R.} \bibnamefont{Kovar}},
  \bibinfo{author}{\bibfnamefont{S.}~\bibnamefont{Banerjee}},
  \bibnamefont{et~al.}, \bibinfo{journal}{Nat. Phys.}
  \textbf{\bibinfo{volume}{15}}, \bibinfo{pages}{696} (\bibinfo{year}{2019}).

\bibitem[{\citenamefont{Jacinto et~al.}(2001)\citenamefont{Jacinto,
  Martinez-Arias, and Martin}}]{AJacinto_NCB_2001}
\bibinfo{author}{\bibfnamefont{A.}~\bibnamefont{Jacinto}},
  \bibinfo{author}{\bibfnamefont{A.}~\bibnamefont{Martinez-Arias}},
  \bibnamefont{and} \bibinfo{author}{\bibfnamefont{P.}~\bibnamefont{Martin}},
  \bibinfo{journal}{Nat. Cell Biol.} \textbf{\bibinfo{volume}{3}},
  \bibinfo{pages}{E117} (\bibinfo{year}{2001}).

\bibitem[{\citenamefont{Martin and Lewis}(1992)}]{PMartin_Nature_1992}
\bibinfo{author}{\bibfnamefont{P.}~\bibnamefont{Martin}} \bibnamefont{and}
  \bibinfo{author}{\bibfnamefont{J.}~\bibnamefont{Lewis}},
  \bibinfo{journal}{Nature} \textbf{\bibinfo{volume}{360}},
  \bibinfo{pages}{179} (\bibinfo{year}{1992}).

\bibitem[{\citenamefont{Tetley et~al.}(2019)\citenamefont{Tetley, Staddon,
  Heller, Hoppe, Banerjee, and Mao}}]{RTetley_NP_2019}
\bibinfo{author}{\bibfnamefont{R.~J.} \bibnamefont{Tetley}},
  \bibinfo{author}{\bibfnamefont{M.~F.} \bibnamefont{Staddon}},
  \bibinfo{author}{\bibfnamefont{D.}~\bibnamefont{Heller}},
  \bibinfo{author}{\bibfnamefont{A.}~\bibnamefont{Hoppe}},
  \bibinfo{author}{\bibfnamefont{S.}~\bibnamefont{Banerjee}}, \bibnamefont{and}
  \bibinfo{author}{\bibfnamefont{Y.}~\bibnamefont{Mao}}, \bibinfo{journal}{Nat.
  Phys.} \textbf{\bibinfo{volume}{15}}, \bibinfo{pages}{1195}
  (\bibinfo{year}{2019}).

\bibitem[{\citenamefont{Brugu{\'{e}}s et~al.}(2014)\citenamefont{Brugu{\'{e}}s,
  Anon, Conte, Veldhuis, Gupta, Colombelli, Mu{\~{n}}oz, Brodland, Ladoux, and
  Trepat}}]{ABrugues_NP_2014}
\bibinfo{author}{\bibfnamefont{A.}~\bibnamefont{Brugu{\'{e}}s}},
  \bibinfo{author}{\bibfnamefont{E.}~\bibnamefont{Anon}},
  \bibinfo{author}{\bibfnamefont{V.}~\bibnamefont{Conte}},
  \bibinfo{author}{\bibfnamefont{J.~H.} \bibnamefont{Veldhuis}},
  \bibinfo{author}{\bibfnamefont{M.}~\bibnamefont{Gupta}},
  \bibinfo{author}{\bibfnamefont{J.}~\bibnamefont{Colombelli}},
  \bibinfo{author}{\bibfnamefont{J.~J.} \bibnamefont{Mu{\~{n}}oz}},
  \bibinfo{author}{\bibfnamefont{G.~W.} \bibnamefont{Brodland}},
  \bibinfo{author}{\bibfnamefont{B.}~\bibnamefont{Ladoux}}, \bibnamefont{and}
  \bibinfo{author}{\bibfnamefont{X.}~\bibnamefont{Trepat}},
  \bibinfo{journal}{Nat. Phys.} \textbf{\bibinfo{volume}{10}},
  \bibinfo{pages}{683} (\bibinfo{year}{2014}).

\bibitem[{\citenamefont{Javierre et~al.}(2009)\citenamefont{Javierre, Vermolen,
  Vuik, and Van~der Zwaag}}]{Javierre_JMB_2009}
\bibinfo{author}{\bibfnamefont{E.}~\bibnamefont{Javierre}},
  \bibinfo{author}{\bibfnamefont{F.}~\bibnamefont{Vermolen}},
  \bibinfo{author}{\bibfnamefont{C.}~\bibnamefont{Vuik}}, \bibnamefont{and}
  \bibinfo{author}{\bibfnamefont{S.}~\bibnamefont{Van~der Zwaag}},
  \bibinfo{journal}{J. Math. Biol.} \textbf{\bibinfo{volume}{59}},
  \bibinfo{pages}{605} (\bibinfo{year}{2009}).

\bibitem[{\citenamefont{Tranquillo and Murray}(1993)}]{RTranquillo_JSR_1993}
\bibinfo{author}{\bibfnamefont{R.~T.} \bibnamefont{Tranquillo}}
  \bibnamefont{and} \bibinfo{author}{\bibfnamefont{J.}~\bibnamefont{Murray}},
  \bibinfo{journal}{J. Surg. Res.} \textbf{\bibinfo{volume}{55}},
  \bibinfo{pages}{233} (\bibinfo{year}{1993}).

\bibitem[{\citenamefont{Sami et~al.}(2019)\citenamefont{Sami, Heiba, and
  Abdellatif}}]{DSami_WM_2019}
\bibinfo{author}{\bibfnamefont{D.~G.} \bibnamefont{Sami}},
  \bibinfo{author}{\bibfnamefont{H.~H.} \bibnamefont{Heiba}}, \bibnamefont{and}
  \bibinfo{author}{\bibfnamefont{A.}~\bibnamefont{Abdellatif}},
  \bibinfo{journal}{Wound Med.} \textbf{\bibinfo{volume}{24}},
  \bibinfo{pages}{8} (\bibinfo{year}{2019}).

\bibitem[{\citenamefont{Eming et~al.}(2017)\citenamefont{Eming, Wynn, and
  Martin}}]{SEming_Science_2017}
\bibinfo{author}{\bibfnamefont{S.~A.} \bibnamefont{Eming}},
  \bibinfo{author}{\bibfnamefont{T.~A.} \bibnamefont{Wynn}}, \bibnamefont{and}
  \bibinfo{author}{\bibfnamefont{P.}~\bibnamefont{Martin}},
  \bibinfo{journal}{Science} \textbf{\bibinfo{volume}{356}},
  \bibinfo{pages}{1026} (\bibinfo{year}{2017}).

\bibitem[{\citenamefont{Purnell and Hines}(2017)}]{BPurnell_Science_2017}
\bibinfo{author}{\bibfnamefont{B.~A.} \bibnamefont{Purnell}} \bibnamefont{and}
  \bibinfo{author}{\bibfnamefont{P.~J.} \bibnamefont{Hines}},
  \bibinfo{journal}{Science} \textbf{\bibinfo{volume}{356}},
  \bibinfo{pages}{1020} (\bibinfo{year}{2017}).

\bibitem[{\citenamefont{Reinke and Sorg}(2012)}]{JMReinke_ESR_2012}
\bibinfo{author}{\bibfnamefont{J.}~\bibnamefont{Reinke}} \bibnamefont{and}
  \bibinfo{author}{\bibfnamefont{H.}~\bibnamefont{Sorg}},
  \bibinfo{journal}{Eur. Surg. Res.} \textbf{\bibinfo{volume}{49}},
  \bibinfo{pages}{35} (\bibinfo{year}{2012}).

\bibitem[{\citenamefont{Huber et~al.}(2013)\citenamefont{Huber, Schnau{\ss},
  R{\"{o}}nicke, Rauch, M{\"{u}}ller, F{\"{u}}tterer, and
  K{\"{a}}s}}]{FHuber_AP_2013}
\bibinfo{author}{\bibfnamefont{F.}~\bibnamefont{Huber}},
  \bibinfo{author}{\bibfnamefont{J.}~\bibnamefont{Schnau{\ss}}},
  \bibinfo{author}{\bibfnamefont{S.}~\bibnamefont{R{\"{o}}nicke}},
  \bibinfo{author}{\bibfnamefont{P.}~\bibnamefont{Rauch}},
  \bibinfo{author}{\bibfnamefont{K.}~\bibnamefont{M{\"{u}}ller}},
  \bibinfo{author}{\bibfnamefont{C.}~\bibnamefont{F{\"{u}}tterer}},
  \bibnamefont{and}
  \bibinfo{author}{\bibfnamefont{J.}~\bibnamefont{K{\"{a}}s}},
  \bibinfo{journal}{Adv. Phys.} \textbf{\bibinfo{volume}{62}},
  \bibinfo{pages}{1} (\bibinfo{year}{2013}).

\bibitem[{\citenamefont{Lee et~al.}(2019)\citenamefont{Lee, Talman, Peirce, and
  Holmes}}]{JLee_BMM_2019}
\bibinfo{author}{\bibfnamefont{J.~J.} \bibnamefont{Lee}},
  \bibinfo{author}{\bibfnamefont{L.}~\bibnamefont{Talman}},
  \bibinfo{author}{\bibfnamefont{S.~M.} \bibnamefont{Peirce}},
  \bibnamefont{and} \bibinfo{author}{\bibfnamefont{J.~W.}
  \bibnamefont{Holmes}}, \bibinfo{journal}{Biomech. Model. Mechanobiol.}
  \textbf{\bibinfo{volume}{18}}, \bibinfo{pages}{1297} (\bibinfo{year}{2019}).

\bibitem[{\citenamefont{Rold{\'{a}}n et~al.}(2019)\citenamefont{Rold{\'{a}}n,
  Mu{\~{n}}oz, and S{\'{a}}ez}}]{LRoldan_CMAME_2019}
\bibinfo{author}{\bibfnamefont{L.}~\bibnamefont{Rold{\'{a}}n}},
  \bibinfo{author}{\bibfnamefont{J.~J.} \bibnamefont{Mu{\~{n}}oz}},
  \bibnamefont{and}
  \bibinfo{author}{\bibfnamefont{P.}~\bibnamefont{S{\'{a}}ez}},
  \bibinfo{journal}{Comput. Methods Appl. Mech. Eng.}
  \textbf{\bibinfo{volume}{350}}, \bibinfo{pages}{28} (\bibinfo{year}{2019}).

\bibitem[{\citenamefont{Guerra et~al.}(2018)\citenamefont{Guerra, Belinha, and
  Jorge}}]{AGuerra_JTB_2018}
\bibinfo{author}{\bibfnamefont{A.}~\bibnamefont{Guerra}},
  \bibinfo{author}{\bibfnamefont{J.}~\bibnamefont{Belinha}}, \bibnamefont{and}
  \bibinfo{author}{\bibfnamefont{R.~N.} \bibnamefont{Jorge}},
  \bibinfo{journal}{J. Theor. Biol.} \textbf{\bibinfo{volume}{459}},
  \bibinfo{pages}{1} (\bibinfo{year}{2018}).

\bibitem[{\citenamefont{Camley and Rappel}(2017)}]{BCamley_JPD_2017}
\bibinfo{author}{\bibfnamefont{B.~A.} \bibnamefont{Camley}} \bibnamefont{and}
  \bibinfo{author}{\bibfnamefont{W.-J.} \bibnamefont{Rappel}},
  \bibinfo{journal}{J. Phys. D: Appl. Phys.} \textbf{\bibinfo{volume}{50}},
  \bibinfo{pages}{113002} (\bibinfo{year}{2017}).

\bibitem[{\citenamefont{Vermolen}(2016)}]{FVermolen_EB_2016}
\bibinfo{author}{\bibfnamefont{F.~J.} \bibnamefont{Vermolen}}, in
  \emph{\bibinfo{booktitle}{Encyclopedia of Cell Biology}}
  (\bibinfo{publisher}{Academic Press}, \bibinfo{year}{2016}),
  vol.~\bibinfo{volume}{4}, p. \bibinfo{pages}{117}.

\bibitem[{\citenamefont{Tartarini and Mele}(2016)}]{DTartarini_FBB_2016}
\bibinfo{author}{\bibfnamefont{D.}~\bibnamefont{Tartarini}} \bibnamefont{and}
  \bibinfo{author}{\bibfnamefont{E.}~\bibnamefont{Mele}},
  \bibinfo{journal}{Front. Bioeng. Biotechnol.} \textbf{\bibinfo{volume}{3}},
  \bibinfo{pages}{206} (\bibinfo{year}{2016}).

\bibitem[{\citenamefont{O'Dea et~al.}(2012)\citenamefont{O'Dea, Byrne, and
  Waters}}]{ODea_MTEB_2012}
\bibinfo{author}{\bibfnamefont{R.}~\bibnamefont{O'Dea}},
  \bibinfo{author}{\bibfnamefont{H.}~\bibnamefont{Byrne}}, \bibnamefont{and}
  \bibinfo{author}{\bibfnamefont{S.}~\bibnamefont{Waters}}, in
  \emph{\bibinfo{booktitle}{Computational Modeling in Tissue Engineering}}
  (\bibinfo{publisher}{Springer}, \bibinfo{year}{2012}),
  vol.~\bibinfo{volume}{10}, p. \bibinfo{pages}{229}.

\bibitem[{\citenamefont{Cumming et~al.}(2010)\citenamefont{Cumming, McElwain,
  and Upton}}]{BCumming_JRSI_2010}
\bibinfo{author}{\bibfnamefont{B.~D.} \bibnamefont{Cumming}},
  \bibinfo{author}{\bibfnamefont{D.}~\bibnamefont{McElwain}}, \bibnamefont{and}
  \bibinfo{author}{\bibfnamefont{Z.}~\bibnamefont{Upton}}, \bibinfo{journal}{J.
  R. Soc. Interface} \textbf{\bibinfo{volume}{7}}, \bibinfo{pages}{19}
  (\bibinfo{year}{2010}).

\bibitem[{\citenamefont{Arciero et~al.}(2011)\citenamefont{Arciero, Mi, Branca,
  Hackam, and Swigon}}]{JArciero_BJ_2011}
\bibinfo{author}{\bibfnamefont{J.~C.} \bibnamefont{Arciero}},
  \bibinfo{author}{\bibfnamefont{Q.}~\bibnamefont{Mi}},
  \bibinfo{author}{\bibfnamefont{M.~F.} \bibnamefont{Branca}},
  \bibinfo{author}{\bibfnamefont{D.~J.} \bibnamefont{Hackam}},
  \bibnamefont{and} \bibinfo{author}{\bibfnamefont{D.}~\bibnamefont{Swigon}},
  \bibinfo{journal}{Biophys. J.} \textbf{\bibinfo{volume}{100}},
  \bibinfo{pages}{535} (\bibinfo{year}{2011}).

\bibitem[{\citenamefont{Arciero et~al.}(2013)\citenamefont{Arciero, Mi, Branca,
  Hackam, and Swigon}}]{JArciero_WRR_2013}
\bibinfo{author}{\bibfnamefont{J.~C.} \bibnamefont{Arciero}},
  \bibinfo{author}{\bibfnamefont{Q.}~\bibnamefont{Mi}},
  \bibinfo{author}{\bibfnamefont{M.}~\bibnamefont{Branca}},
  \bibinfo{author}{\bibfnamefont{D.}~\bibnamefont{Hackam}}, \bibnamefont{and}
  \bibinfo{author}{\bibfnamefont{D.}~\bibnamefont{Swigon}},
  \bibinfo{journal}{Wound Repair Regen} \textbf{\bibinfo{volume}{21}},
  \bibinfo{pages}{256} (\bibinfo{year}{2013}).

\bibitem[{\citenamefont{Geris et~al.}(2013)}]{LGeris_book_2013}
\bibinfo{author}{\bibfnamefont{L.}~\bibnamefont{Geris}} \bibnamefont{et~al.},
  \emph{\bibinfo{title}{Computational modeling in tissue engineering}}
  (\bibinfo{publisher}{Springer}, \bibinfo{year}{2013}).

\bibitem[{\citenamefont{Carvalho et~al.}(2018)\citenamefont{Carvalho, Patricio,
  Ponte, Heisenberg, Almeida, Nunes, Ara{\'{u}}jo, and
  Jacinto}}]{LCarvalho_JCB_2018}
\bibinfo{author}{\bibfnamefont{L.}~\bibnamefont{Carvalho}},
  \bibinfo{author}{\bibfnamefont{P.}~\bibnamefont{Patricio}},
  \bibinfo{author}{\bibfnamefont{S.}~\bibnamefont{Ponte}},
  \bibinfo{author}{\bibfnamefont{C.~P.} \bibnamefont{Heisenberg}},
  \bibinfo{author}{\bibfnamefont{L.}~\bibnamefont{Almeida}},
  \bibinfo{author}{\bibfnamefont{A.~S.} \bibnamefont{Nunes}},
  \bibinfo{author}{\bibfnamefont{N.~A.~M.} \bibnamefont{Ara{\'{u}}jo}},
  \bibnamefont{and} \bibinfo{author}{\bibfnamefont{A.}~\bibnamefont{Jacinto}},
  \bibinfo{journal}{J. Cell Biol.} \textbf{\bibinfo{volume}{217}},
  \bibinfo{pages}{4267} (\bibinfo{year}{2018}).

\bibitem[{\citenamefont{Bonnet et~al.}(2012)\citenamefont{Bonnet, Marcq,
  Bosveld, Fetler, Bella{\"\i}che, and Graner}}]{IBonnet_JRSI_2012}
\bibinfo{author}{\bibfnamefont{I.}~\bibnamefont{Bonnet}},
  \bibinfo{author}{\bibfnamefont{P.}~\bibnamefont{Marcq}},
  \bibinfo{author}{\bibfnamefont{F.}~\bibnamefont{Bosveld}},
  \bibinfo{author}{\bibfnamefont{L.}~\bibnamefont{Fetler}},
  \bibinfo{author}{\bibfnamefont{Y.}~\bibnamefont{Bella{\"\i}che}},
  \bibnamefont{and} \bibinfo{author}{\bibfnamefont{F.}~\bibnamefont{Graner}},
  \bibinfo{journal}{J. R. Soc. Interface} \textbf{\bibinfo{volume}{9}},
  \bibinfo{pages}{2614} (\bibinfo{year}{2012}).

\bibitem[{\citenamefont{Tlili et~al.}(2015)\citenamefont{Tlili, Gay, Graner,
  Marcq, Molino, and Saramito}}]{STlili_EPJE_2015}
\bibinfo{author}{\bibfnamefont{S.}~\bibnamefont{Tlili}},
  \bibinfo{author}{\bibfnamefont{C.}~\bibnamefont{Gay}},
  \bibinfo{author}{\bibfnamefont{F.}~\bibnamefont{Graner}},
  \bibinfo{author}{\bibfnamefont{P.}~\bibnamefont{Marcq}},
  \bibinfo{author}{\bibfnamefont{F.}~\bibnamefont{Molino}}, \bibnamefont{and}
  \bibinfo{author}{\bibfnamefont{P.}~\bibnamefont{Saramito}},
  \bibinfo{journal}{Eur. Phys. J. E} \textbf{\bibinfo{volume}{38}},
  \bibinfo{pages}{1} (\bibinfo{year}{2015}).

\bibitem[{\citenamefont{Dill}(2007)}]{EDill_CRC_2007}
\bibinfo{author}{\bibfnamefont{E.~H.} \bibnamefont{Dill}},
  \emph{\bibinfo{title}{Continuum mechanics: elasticity, plasticity,
  viscoelasticity}} (\bibinfo{publisher}{CRC press}, \bibinfo{year}{2007}).

\bibitem[{\citenamefont{Landau and Lifchitz}(1990)}]{LLandau_EM_1990}
\bibinfo{author}{\bibfnamefont{L.}~\bibnamefont{Landau}} \bibnamefont{and}
  \bibinfo{author}{\bibfnamefont{E.}~\bibnamefont{Lifchitz}},
  \emph{\bibinfo{title}{Physique Th{\'e}orique 7: Th{\'e}orie de
  l'{\'e}lasticit{\'e}}} (\bibinfo{publisher}{Ed. Mir}, \bibinfo{year}{1990}).

\end{thebibliography}
\end{document}